\documentclass[twocolumn]{emulateapj}
\usepackage{graphicx,url,longtable,natbib}
\newcommand{\kms}{\>{\rm km}\,{\rm s}^{-1}}
\newcommand\Log{\,{\rm Log}\,}
\newcommand\Vmax{V_{\rm max}}

\shorttitle{Morphological dependent TFR}
\shortauthors{Shen et al.}
\begin{document}

\title{The morphological dependent Tully-Fisher relation of spiral galaxies}
\author{Shiyin Shen\altaffilmark{1,3}, Caihong Wang\altaffilmark{1,2}, Ruixiang Chang\altaffilmark{1,3},
 Zhengyi Shao\altaffilmark{1,3}, Jinliang Hou\altaffilmark{1,3}, Chenggang Shu\altaffilmark{3}}
 \affil{$^1$ Key Laboratory for Research in Galaxies and Cosmology, Shanghai Astronomical
Observatory, Chinese Academy of Sciences, 80 Nandan Road, Shanghai, 200030,
China}
 \email{ssy@shao.ac.cn}
 \affil{$^2$ Graduate University of Chinese Academy of Sciences}
 \affil{$^3$ Key Lab for Astrophysics, Shanghai 200234}

\begin{abstract}
The Tully-Fisher relation of spiral galaxies shows notable dependence on
morphological types, with earlier type spirals having systematically lower
luminosity at fixed maximum rotation velocity $\Vmax$. This decrement of
luminosity is more significant in shorter wavelengths. By modeling the rotation
curve and stellar population of different morphological type spiral galaxies in
combination, we find the $\Vmax$ of spiral galaxies is weakly dependent on the
morphological type, whereas the difference of the stellar population
originating from the bulge disk composition effect mainly account for the
morphological type dependence of the Tully-Fisher relation.

\end{abstract}

\keywords{galaxies: spiral - galaxies: stellar content - galaxies: kinematics and dynamics}

\maketitle
\section{introduction}
The Tully-Fisher relation \citep[][hereafter TFR]{Tully77} is an empirical
relation between  the absolute magnitude $M$ and the maximum rotation velocity
$\Vmax$ of spiral galaxies, which  is typically expressed as
\begin{equation}
M=\alpha \Log \left(\frac{\Vmax}{200\kms}\right) + \beta\,,
\end{equation}
where the values of slope $\alpha$ and  zero-point $\beta$ are dependent on the
photometric band of the $M$ being measured.

It has long been known that the TFR has a morphological type dependence, with
earlier type spirals having systematically lower luminosity at fixed $\Vmax$
\citep{Roberts78,Rubin80}. This luminosity difference is also waveband
dependent, with larger offsets in shorter wavelengths. For example, in $I$
band, \citet{Giovanelli97} found a 0.32 mag lower zero-point of the TFR for
Sa/Sab galaxies and 0.10 mag lower for Sb galaxies than the Sbc and later type
spirals. In $B$ band, \citet{Russell04} found  that Sb galaxies have a
zero-point of 0.57 mag lower than Sc galaxies. In a recent study,
\cite{Russell08} compared the TFR-derived distances of nearby groups and
clusters and found a mean difference of 0.19 mag in $H$ band between the Sb and
Sc spiral galaxies. Moreover, using a large sample of spiral galaxies,
\cite{Masters06,Masters08} found that the morphological dependence of the TFR
is not only a shift of the zero-point, but even dependent on the luminosity in
the way that the differences are more pronounced for more luminous galaxies.

The morphological dependence of the TFR originates from either the differences
of  the stellar population($M$) or the disk dynamics ($\Vmax$) or both.

On  one hand, it is well known that the colors of earlier type spirals are
redder. \citet{Devereux91} interpreted this color difference as originating
from the bulge disk composition effect. The stellar population of bulges are
typically older and more metal rich than that of the disks. A larger fraction
of the bulge component in earlier type spirals naturally results in an redder
color on average for the whole galaxy. However, with a bulge disk decomposed
sample,  \citet{Kennicutt94} studied and compared the star formation histories
of only the disk component of different type spirals and found that the stellar
population of the  disks of later type spiral is also on average younger.

On the other hand, although the rotation curves of spiral galaxies are proposed
to follow a universal shape \citep{Persic91,Persic96}, there is evidence
showing that the rotation curves of early type spirals  rise more rapidly  in
the inner region than that of late types \citep{Corradi90,Noordermeer07a}.
\citet{Noordermeer07b} show that the massive Sa galaxies lie better in the
well-defined TFR when using the asymptotic rotation velocity $V_{\rm{asymp}}$
instead of  $\Vmax$, implying a dependence of $\Vmax$ on the morphological
type.

On the theoretical side, the zero-point, slope and scatter of the observed TFR
can all be well accommodated by the current disk formation model in the
framework of the cold dark matter hierarchical cosmogonies
\citep{Dalcanton97,MMW,Mo00,Pizagno05}. However, the morphological dependence
of the TFRs has not been probed in these studies  because the bulge component
in these models is typically neglected. Another limitation of these models is
that the stellar populations have not been tackled  with a physical
prescription, but with pre-determined mass-to-light ratios.

In this study, we aimed to model the dynamics and stellar population of
different type spiral galaxies in combination  and try to find out which factor
is the main contributor to the morphological dependence of the TFR. In
specific, we will follow  the dynamical  model of Mo, Mao \& White (1998,
hereafter MMW) and extend it to include a bulge component. For the stellar
population, we will parameterize the star formation histories of the disks and
bulges separately and then derive their mass-to-light  ratios in different
bands using the stellar population synthesis code\citep{BC03}.

This paper is organized as follows. In Section 2, we describe our dynamical
model of the spiral galaxies, including the disk, bulge and dark halo
components respectively. In Section 3, we study the stellar properties of the
disks and bulges with parameterized star formation histories. We compare our
model predictions with the observational data in Section 4. We make discussions
on the uncertainties of our results in Section 5 and finally give a brief summary
in Section 6.

\section{dynamical model of spiral galaxies}

Our modeling of the dynamics of spiral galaxies follows and  simplifies the
disk formation model of MMW, but with more attention on the bulge contribution.
The interested reader is referred to MMW for detail. Here we repeat the
essentials related to our study. In this model,  initially, the gas and dark
matter are uniformly mixed in a virialized halo. As a result of  dissipative
and radiative cooling, the gas gradually cools down and settles into a disc
structure due to the conservation of angular momentum.

We define that the stellar mass of the galaxy finally formed is $M_*$ and the
fraction of this mass to  the initial halo mass $M_h$ is $m_s\equiv M_*/M_h$.
We express the bulge fraction of the formed galaxy being $f_b$, so that the
masses of the bulge and disk are $M_b=f_b M_*$ and $M_d=(1-f_b) M_*$
respectively.

The rotation curve of a spiral galaxy is contributed by  three dynamical
components: halo,  bulge and disk. We show and discuss our assumptions on each
term below.

\subsection{halo}
The $N$-body simulations show that the collapsed and virialized dark halos
follow a universal density NFW profile \citep*{NFW96},
\begin{equation}\label{NFW}
  \rho(r) = \frac{V_h^2}{4\pi Gr^2}\frac{c}{[{\rm{ln}}(1+c)-c/(1+c)]}\frac{r/r_s}{(r/r_s+1)^2},
\end{equation}
where  $r_s$ is a scale radius, $G$ is the gravitational constant, $V_h$ is the
circular velocity and $c$ is the concentration parameter. The concentration $c$
is defined as $c\equiv r_{200}/r_s$, where $r_{200}$ is the virial radius
\footnote{Here, $r_{200}$ is the radius within which the mean mass density is
200 times the cosmic critical density $\rho_{crit}$.} of the halo. For a given
cosmology, the viral radius $r_{200}$, halo mass $M_h$ and circular velocity
$V_h$ are related by

\begin{equation}
r_{200}=\frac{V_h}{10H(z)}, ~~~~~~ M_h=\frac{V_h^3}{10GH(z)}\,,
\end{equation}
where $H(z)$ is the Hubble constant at redshift $z$.

In this study, we use the concordance $\Lambda$CDM cosmology model, with
$H_0=70\kms{\rm{Kpc^{-1}}}$, $\Omega_0=0.3$,  $\Omega_{\Lambda}=0.7$ and baryon
mass density $\Omega_B=0.04$.

With the assembling of the baryons into disk and bulge, the gravitational
effect from the bulge and disk changes the initial halo mass distribution
through contraction. We follow MMW and use the adiabatic contraction
assumption to analysis this effect (see MMW for detail).

For a halo with given circular velocity $V_h$, the concentration $c$ is the
only parameter to be quantified. At given redshift $z$, the concentration
parameter $c$ is mainly correlated with the mass of the halos
\citep{NFW96,Bullock01}. We  adopt a simple parametrization of the
concentration $c$ at redshift zero as that in \cite{Shen02}:

\begin{equation} \label{Par_c}
c=8.5\left(\frac{V_h}{100\kms}\right)^{-1/3}
\end{equation}

\subsection{bulge}\label{bulge}

We assume that the bulge component of spiral galaxies has  a spherical mass
distribution and the mass density profile $\rho_b(r)$ follows a $Hernquist$
profile \citep{Hernquist90}, whose projection approximates the classical
$R^{1/4}$ surface brightness profile of elliptical galaxies. The $\rho_b(r)$ is
expressed as
\begin{equation}
\rho_b(r)=\frac{M_b}{2\pi}\frac{a}{r}\frac{1}{(r+a)^3} \,,
\end{equation}
where  $M_b$ is the total bulge  mass, $a$ is the bulge scale radius, which is
correlated with the effective(half-light) radius $R_e$ in the way
$R_e\approx1.82a$[Equ. 38 of \citet{Hernquist90}]. To establish the dynamics of
a bulge with mass $M_b$,  we need to know the scale radius $a$ or effective
radius $R_e$. We assume that the bulges follow the observed size-mass($R-M$)
relation of elliptical galaxies \citep{Shen03},
\begin{equation}
R_e({\rm {Kpc}})=3.47\times10^{-5}\left(\frac{M_b}{M_\odot}\right)^{0.56}\,.
\end{equation}

\subsection{disk}

The surface brightness profile of spiral disks typically follow  an exponential
profile. Here, we assume the disk surface mass density profile is also
exponential,
\begin{equation}
\mu_d(r)=\mu_0 {\rm{exp}}(-r/R_d)
\end{equation}
where $\mu_0$ is the central  surface mass density and $R_d$ is the
scale-length. $\mu_0$ and $R_d$ are related to the total mass of the disk $M_d$
through $M_d=2\pi \mu_0 R_d^2$.

Since the observed size  of the disk in a spiral galaxy is affected by its
central bulge, we do not take the observed $R-M$ relation of spiral galaxies as
that in the model. Following MMW, we assume that the disk size $R_d$ is
determined by its initial angular momentum $J$, which can be parameterized by a
spin parameter $\lambda$
\begin{equation}
\lambda=J|E|^{1/2}G^{-1}M^{-5/2},
\end{equation}
where $E$ is the total energy of the halo. With reasonable assumptions that the
specific angular momentum per particle of the baryons is the same as the dark
matter and there is no angular momentum transfer among different components,
the angular momentum of the disk finally formed is therefore $J_d=m_d J$. Here,
$m_d$ is the fraction of baryons settling into the disk and is equal to
$m_s(1-f_b)$. The scale-length of the disk $R_d$ then equals
 \begin{equation}
R_d=\frac{1}{\sqrt{2}}\lambda r_{200}f^{-0.5}_cf_R(\lambda,c,m_d)\,
 \end{equation}
where $f_c$ is a factor coming from the halo density profile and  only
dependent on concentration $c$ (see Equ. 23 of MMW), $f_R$ is a factor coming
from both the halo density profile and the gravitational effects of disk and
bulge(see Equ. 29 of MMW). It is shown that the size of spiral galaxies
predicted from $\lambda$ is in excellent agreement with the
observations\citep{Syer99,Shen03}.

Numerical simulations have found that the spin parameter $\lambda$ of the dark
matter halo follows a log-normal distribution with median $\bar{\lambda}\sim
0.04$ and scatter $\sigma_{{\rm{ln}}\lambda}\sim0.4$ and this distribution is
quite independent of the cosmology and halo properties\citep{Bett07}.  For the
purpose of simplicity, we do not consider the scatter of the angular momentum
distribution and take the median value $\lambda=0.04$ for all disks(see more
discussions in Section \ref{dis_Lambda}).

\subsection{rotation curve}\label{sec_RC}

With the mass distribution of the bulge, halo and disk components determined
through the model parameters shown above, the rotation curve of a spiral galaxy can
be determined by the sum in quadrature of contributions from these three terms:
\begin{equation}\label{RCequ}
V^2(r) = G\frac{M_{DM}(r)+M_b(r)}{r}+V^2_{d}(r) \,.
\end{equation}
During the calculation of the term $V^2_d$ contributed from the disk
component, the flattened geometry needs to be taken into account \citep{Binney87}.

To fully determine the rotation curve of a given spiral galaxy with  stellar
mass $M_*$, we still need to know the relative proportions  of these three
components (bulge, disk and dark halo). This could be done by settling the two
other parameters, i.e. the baryon fraction $m_s$ and the bulge-disk-ratio
$B/D$\footnote{$B/D\equiv f_b/(1-f_b)$}. We set the baryon fraction $m_s$ to be
a function of the halo mass (in terms of circular velocity
$V_h$)\citep{Shen03},
\begin{equation}\label{ms}
 m_s=\frac{0.13}{1+(V_h/150\kms)^{-2}}\,.
\end{equation}
This assumption is based on the consideration that  gas outflow process in low
mass halo is efficient, due to their shallow potential well, and a large
fraction of baryons is blown out of the dark halo, while the baryon fraction of
high mass halos gradually approach the cosmic baryon fraction
$\Omega_B/\Omega_0\approx0.13$\citep{White91}. The bulge-disk-ratio $B/D$ is
taken to be 0.5/0.3/0.1 for model Sa/Sb/Sc galaxies
respectively\citep{Simien86}.

\begin{figure}
\includegraphics[width=84mm]{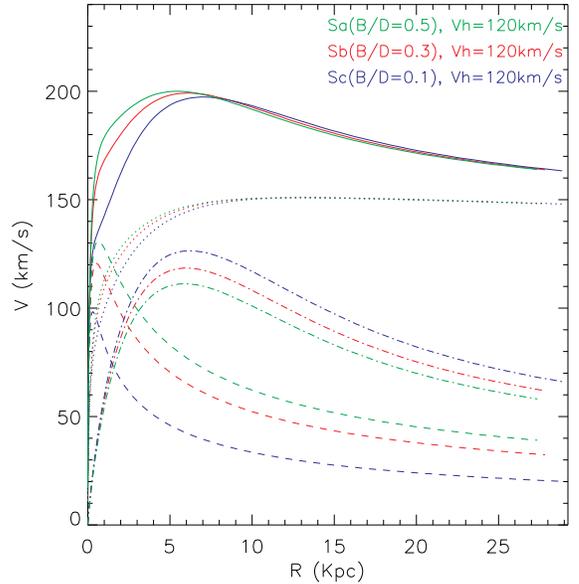}
\caption[Rotation curve]{The rotation curves of  the model galaxies with $V_h=120\kms$.
The solid lines show the rotation curve while the dot-dashed, dotted and  dashed
lines show the contributions from disk, halo and bulge components
respectively. The green, red and blue lines represent the model Sa, Sb and
Sc galaxies respectively.}\label{RC}
\end{figure}

We show three examples of resulting rotation curves for three morphological
types(Sa/Sb/Sc) in Fig. \ref{RC}. All three model galaxies are chosen to have
the same circular velocity $V_h=120\kms$ and thus the same stellar mass
$M=3\times10^{10}M_\odot$. For the earlier type spiral, as we can see from the
figure, due to the larger bulge component, our model predicts a steeper
increase of rotation velocity in the inner region, which brings on the
appearance of  $\Vmax$ at smaller radius, and a systematically larger $\Vmax$.

However, the value of $\Vmax$ only weakly  correlates with the model type. The
$\Vmax$ of the Sa(Sb) spiral is only 1.5(1.0) percent higher than  the
corresponding Sc spiral. This is because  the relatively smaller bulge
fraction of the later type spiral is always compensated by its larger disk
fraction (see the components of rotation curves in Fig. \ref{RC}).

\section{stellar populations of spiral galaxies}
Besides the different dynamics, the bulge and disk also show different stellar
populations. The bulges are typically old and show little recent star formation,
while the star formation time-scales of the disks are long and there is still
ongoing star formation throughout the disk at the present day.

In this study,  we parameter the stellar population of the bulge components
with a single stellar population with age of 10G year(Gy). For the disks,
following the usual convention, we parameterize their star formation
histories(SFH) with an exponential function
\begin{equation} \label{SFH}
SFR(t)\propto {\rm{exp}}(-t/\tau)
\end{equation}
where $\tau$ is the time scale of the SFH. This simple analytical expression
can parameterize different kinds of SFHs.  For $\tau$ approaching 0, Equ.
(\ref{SFH}) represents a single stellar population. When $\tau$ tends to
infinity, the star formation rate is a constant along the history. The age of
the disks is also set to 10\,Gy.

Kennicutt et al. (1994, hereafter K94) parameterized the SFH of the spiral
disks with a $b$ parameter, which is defined as
\begin{equation}
b\equiv \frac{SFR}{<SFR>}
\end{equation}
where $SFR$ is the star formation rate today and $<SFR>$ is  the average star
formation rate in past. They estimated the $b$ parameter from $UBV$ colors and
$H\alpha$ fluxes for different morphological type spirals and found that the
Sa/Sb/Sc disks have typical $b\approx 0.12/0.33/0.84$ and variation of about
$0.05/0.15/0.20$ respectively.\footnote{The $b$ values are taken from Table 4
of K94, where the sub-types include Sa/Sab/Sb/Sbc/Sc/Scd spirals.  We average
the $b$ values of Sa and Sab sub-types as the Sa type. The $b$ value of Sb type
is kept. For Sc type, we average the $b$ values of Sbc and Sc/Scd sub-types.
The variations of $b$ are estimated from Fig.6 of K94.} With the SFHs
parameterized by  Equ. \ref{SFH} as a prior,  these $b$ values correspond to
the $\tau$ with median 3/5/30\,Gy and $1\sigma$ range [2,4]/[4,7]/[10,$\infty$]
Gy for Sa/Sb/Sc spirals respectively. \footnote{Here, the $1\sigma$ range
denotes the 32th to 68th percentiles of the $\tau$ distribution and is
calculated from the estimated variation of $b$. For Sc spirals, during the
calculation of the range of $\tau$, we force the upper limit of $b$ to be 1,
corresponding to $\tau \sim \infty$, which gives the youngest average stellar
population  as that can do by Equ. \ref{SFH}.}

Based on the SFHs assumed for disk and bulge components, we use the new version
of the stellar population synthesis code of \cite{BC03}, denoted as CB07, to
calculate their mass-to-light ratios in different wavebands. The metalicity of
the stellar population is set to be the solar value while the stellar initial
mass function is taken from \cite{Chabrier03} with lower and upper mass limit
being $0.1M_\odot$ and $100M_\odot$ respectively.

Finally, with the predetermined stellar mass $m_s$ and $B/D$ for each model
galaxy, we calculate its absolute magnitudes in different optical wavebands.

\section{model predictions}
In this section, we show our model predicted  morphologically dependent TFRs
and compare them with the observational results. For each morphological type,
we build 11 model galaxies with halo circular velocity $V_h$ in the range from
100 to $200\kms$ with an interval of $10\kms$. With the baryon fraction
characterized by Equ. \ref{ms}, the range of the stellar mass of our model
galaxies is thus from $1.4\times10^{10}$  to $2.3\times10^{11}M_\odot$.

In our model, there is only one free  parameter, i.e. the disk star formation
time scale $\tau$ in Equ. \ref{SFH}. We first set the $\tau$ values of
different type spirals to be these suggested by K94, i.e. the median
$\tau=3/5/30$\,Gy and its 1$\sigma$ range being [2,4]/[4,7]/[10,$\infty$] Gy
for Sa/Sb/Sc spirals respectively. We refer to this model setting as K94 model
below.

\subsection{$I$-band TFRs}\label{ITF}

\begin{figure}
\includegraphics[width=84mm]{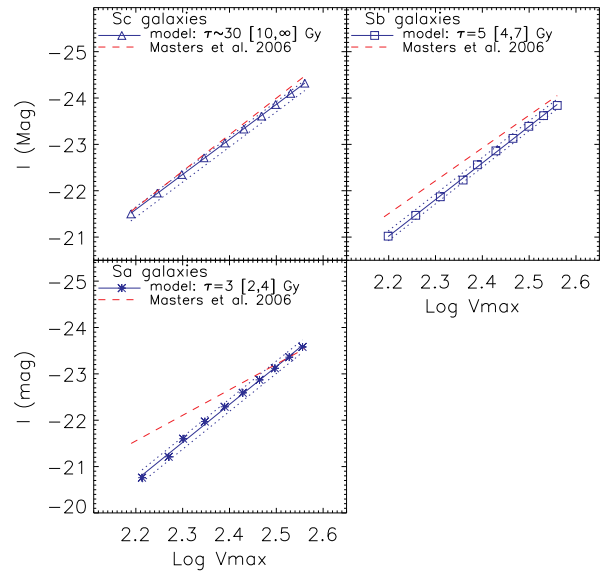}
\caption[TFR spirals]{The model predicted $I$ band TFRs of different morphological type
spiral galaxies. The triangles,squares and stars represent the model Sc, Sb and
Sa  galaxies in top left, top right and bottom left panels respectively. The
solid lines show the best TFR fittings of model galaxies while the observed
TFRs of M06 are shown as the  dashed lines for comparison. The dotted lines in
each panel show the range of the model predicted TFRs when the star formation
time scale $\tau$ varies in its 1$\sigma$ range (as that denoted in the
brackets).   }

\label{TFR}
\end{figure}
Lots of recent observational TFR studies are  in the near-infrared $I$
band\citep[e.g.][]{Giovanelli97,Dale99,Vogt04}, which takes the advantage of
better detector photometry than $H$ band and less scatter in TFR than blue $B$ band.
More recently, \cite{Springob07} compiled and published a new generation of
a homogeneous  galaxy catalog, referred to as the {\it SFI++}, which contains $\sim$
5000 spirals suitable for TFR study in $I$ band. Based on the {\it SFI++}
catalog, the TFRs of Sa, Sb and Sc type spirals have been presented by
\citet[][hereafter M06]{Masters06}. Here, we first show our model predicted $I$
band TFRs for different type galaxies and make comparisons with the results of
M06.

The K94 model predicted  $I$ band TFRs of Sc, Sb and Sa spirals are shown in
the top left, top right and bottom left panels of Fig. \ref{TFR} respectively.
The model galaxies are represented by the triangles(Sc), squares(Sb) and
stars(Sa) while the fitted TFRs are shown as the solid lines in each panel. The
observed $I$ band TFRs of different type spirals of M06 are shown as the dashed
lines for comparison. The dotted lines in each panel show the ranges of model
predicted TFRs when the disk star formation time scale $\tau$ varies in its
1$\sigma$ range.  Inside its range, a larger $\tau$ results a larger zero-point
of the TFR, i.e. galaxies will be brighter at given $\Vmax$ for larger $\tau$.

Sc type spirals(include Sbc/Sc/Scd) are the dominant morphological type in TFR
studies\footnote{The later type spirals are more likely to be strong HI or
H$\alpha$ emitters so that the measurement of the rotation width is easier.}. Lots
of the TFR studies use Sc spirals as the reference and make morphological type
corrections on other types, e.g. \cite{Giovanelli97, Dale99, Ziegler02}. For Sc
spirals, as we can see, by setting the model parameters to be either the
typical values from numerical simulations(e.g. $\lambda$ and $c$) or being
constrained from other observations(e.g. $\tau\sim$ 30\,Gy) without any further
fine-tuning, our model predicted $I$ band TFR is in excellent agreement with
the observations.

However, with default settings of model parameters, our model predicted TFRs of
Sb and Sa spirals are not well consistent with the observations of M06. The
predicted TFR of Sb spirals($\tau$=5\,Gy) have a systematically larger
zero-point than observations, i.e. the model predicted luminosity is too
bright. When $\tau$ reaches its upper range 7 Gy, the discrepancy becomes
smaller, but still be far from consistent with observations. We will discuss
this discrepancy in more detail in next section.

For Sa spirals, the slope of the observed TFR is significantly shallower than
our model predictions. However, as that discussed in \cite{Giovanelli97} and
M06, the disagreement of the TFR slope of the Sa spirals should be interpreted
in caution because the incompleteness of the Sa spirals is large, especially in
the low luminosity end, which will artificially bias the slope to lower value.
On the other hand, our modeling on the $B/D $ of Sa spirals is a constant 0.5,
which might be too simplified. In the study of M06, all the S0/Sa/Sab galaxies
are grouped as Sa type. Among three sub-types, the earlier type galaxies, e.g.
S0, having systematically larger $B/D$, are also systematically brighter.
Therefore, from low $\Vmax$ to high $\Vmax$ galaxies, there is actually a
systematical change of $B/D$, which will also shallow the slope of TFR. By
introducing a systematical change of $B/D$ as function of $\Vmax$, we can in
principle reproduce the slope of Sa spirals as observed. However, we prefer not
to tune our model to reproduce this relation with a cost of introducing more
uncertainties.

Comparing with the uncertainties in the modeling  of the TFR slope, the
zero-point is quite robust.  Moreover, many studies of  the morphological
dependence of the TFRs only report a global shift of the zero-point, e.g.
\citet{Giovanelli97,Russell04,Russell08} (see next section). Therefore, in the
following section, we parameterize the morphological dependence of the TFRs
only with the shift of the zero-point $\Delta M$.

\subsection{morphology dependent TFRs in different bands}

We choose the zero-point shift $\Delta M$ of the TFRs at $\Vmax =200\kms$,
where the corresponding absolute magnitude is roughly the $M_*$ of the
luminosity function of spiral galaxies and where the number counts of galaxies
in a flux-limited sample normally peaks. In this case, $\Delta M=\Delta \beta$
of Equ. 1.

\begin{figure}
\includegraphics[width=84mm]{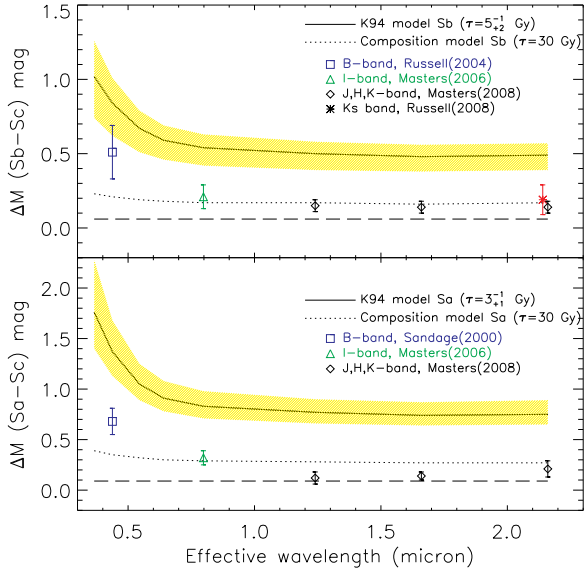}
\caption[morphological dependent TFR]{The morphology dependence of the TFR in different
wavebands. The solid and dotted curves show the predicted shift of the
zero-point of the TFRs, $\Delta M$, as function of wavelength from the K94
model and `composition' model respectively(see text). The dashed horizontal
lines show the contributions to $\Delta M$ from the  $\Vmax$(see text). The
upper panel shows $\Delta M$ between Sb and Sc spirals, while the $\Delta M$
between Sa and Sc is shown in the bottom panel. The observed $\Delta M$ in
$B,I,J,H$ and $K$ band are labeled for comparison. } \label{ZP_TFR}
\end{figure}

We show the $\Delta M$ between  Sb and Sc spirals in the top panel of Fig.
\ref{ZP_TFR}, whereas the  $\Delta M$  between Sa and Sc spirals is shown in
the bottom panel. The model predicted $\Delta M$ in different wavebands are
connected  as a function of their effective wavelengths with lines. The
wavebands include  $U,B,V,R,I,J,H,K$ and the wavelength ranges from 0.33 to 2.2
micron. The observed difference of $\Delta M$ at $\Vmax=200\kms$ in different
bands found in the literature are plotted against their effective wavelengths.
Results from different authors are labeled with different symbol types. The
references and details of these morphology dependent TFRs are listed in Table
1. A few other studies in the literature are not quoted in Table 1 and Fig.
\ref{ZP_TFR}, including \cite{Rubin85}, \cite{Giovanelli97} and the TFR of Sb
spirals of \cite{Sandage00}. \cite{Rubin85} found the magnitude differences
$\Delta M$ between Sa and Sc spirals are as high as 2 mag in $B$ band and 1 mag
in $H$ band, whose number of each type galaxies is limited($<20$) and  without
error estimation quoted for the parameters of TFRs. The $I$ band data of
\cite{Giovanelli97} has been expanded and re-analyzed by M06 and their results
are generally consistent. The TFR of Sb spirals in \cite{Sandage00} is
arbitrarily neglected, where its zero-point is even lower than Sc spirals, i.e.
$\Delta M$(Sb-Sc)$<0$, contradictory to most of the other studies.

The solid lines in Fig. \ref{ZP_TFR} show  predictions from the K94 model, i.e.
the model with $\tau=3,5,30$\,Gy for Sa,Sb,Sc spirals respectively. The
shadowed regions show the ranges of model predicted $\Delta M$ when $\tau$
varies in its 1$\sigma$ range. As we can see, the K94 model over-predicts the
differences of the zero-points significantly for both Sb/Sc and Sa/Sc pairs.
The average observed zero-point differences in near-infrared bands are $\Delta
M$(Sb-Sc)$\sim$0.15\,mag and $\Delta M$(Sa-Sc)$\sim$0.20\,mag, whereas the
model prediction is as high as $\Delta M$(Sb-Sc)$\sim$0.45\,mag  and $\Delta
M$(Sa-Sc)$\sim$0.60\,mag. That means, if the differences of the stellar
populations of the disks of different type spirals are as that suggested by
K94, the shift of the zero-point of the TFRs would be much larger than
observed.

With the disk star formation time scale $\tau$ approaching  its upper range(the
lower boundary of the shadowed region), the discrepancy between the model
predicted $\Delta M$ and the observations becomes  smaller. This result prompts
us to consider an extreme case, $\tau=30$Gy for both Sa and Sb disks. In this
case, the stellar populations of the disks of different type spirals are the
same, while the different global colors of different type spirals only
originate from  the bulge disk composition effect. This scenario is consistent
with the results of \citet{Devereux91}. To distinguish it from the K94 model,
we refer to this scenario($\tau=30$ Gy for all type disks) as the `composition'
model below.

For the `composition' model, the model predicted $\Delta M$ as function of the
effective wavelength is shown as the dotted lines in Fig. \ref{ZP_TFR}.
Surprisingly, the predicted $\Delta M$ from this model, especially in
near-infrared bands, is generally consistent with observations, either for
Sb/Sc or Sa/Sc pairs. The predicted $\Delta M$ in $B$ band is smaller than that
observed. A possible solution to this discrepancy is the dust extinction which
has not been taken into account in our model. If the face-on dust extinction of
earlier type spirals is more significant, the $\Delta M$ in blue band will be
larger than now we predicted(see detailed discussions in  Section \ref{dust}).

Finally, it is worth mentioning that the $\Delta M$ between different
spiral types is not contributed by the stellar population alone but composed by
both the dynamics and stellar population. As we have shown in section
\ref{sec_RC}, for the galaxies with the same stellar mass
$3\times10^{10}M_\odot$, the $\Vmax$ of a Sa(Sb) spiral is about 1.5(1.0)
percent higher than  Sc spiral, which corresponds to the $\Delta M$
$\sim0.04(0.03)$\,mag, independent of the wavebands. This contribution is shown
as two dashed horizontal  lines in the top and bottom panels of Fig.
\ref{ZP_TFR}. However, as we can see, the stellar population is still the
dominant contributor to the morphological dependence of the TFR.

\section{ discussion}

There are some uncertainties in our modeling of both dynamics and stellar
population of the spiral galaxies, e.g. the scatters of model parameter, the
diverse bulge properties and the internal dust extinction etc. We discuss these
issues below.

\subsection{scatters of model parameter}\label{dis_Lambda}

In our modeling, we only take the typical values for each model parameter and
do not consider their scatters  except the key parameter $\tau$. However, all
these model parameters  show  scatters as suggested by either  numeric
simulations or observations, e.g. \citet{Mo00,Jing00,Shen02}. If the scatters
of model parameters are independent variables and do not correlate with the
morphological types(as suggested in our model), these scatters will only make
contributions to the scatters of the predicted TFRs and will not affect any of
our conclusions \citep{Mo00,Shen02}.

However, the morphological type may correlate with the spin value of a galaxy.
A smaller $\lambda$ may trigger a formation of larger bulge easier through disk
instability \citep{Shen03}, so that induce an earlier type spiral. In other
words, the earlier type spirals would be biased to the galaxies with
systematically lower $\lambda$ when we considered the scatter of $\lambda$. In
this case, the resulted $\Vmax$ of earlier type spirals will be even larger
because of the lower $\lambda$\citep{Mo00} and thus the room left for the
stellar population in $\Delta M$ is less.

\subsection{pseudo and classical bulges}

In this study, we have treated all the bulges as scaled elliptical, i.e. the
classical bugles, for simplicity.

The current view on the bulges is quite complex. There are mainly two types of
bulge, i.e. the classical bulge and the pseudo-bulge, which may be originated
from different physical formation processes\citep{Fisher08,Fisher09,Gadotti09}.
The classical bulges are similar to scaled elliptical galaxies, with their
surface brightness following classical $R^{1/4}$ profile and  colors being red.
The pseudo-bulges tend to show younger stellar population and their density
profiles can be characterized by a S\'ersic profile with S\'ersic index $n$
significantly smaller than 4. The classical bulges typically appear in earlier
type spirals and their masses are high. The majority of the stellar mass of the
bulges is in classical bulges [$\sim$ 90\%, \citet{Gadotti09}].

If the bulges of late type(e.g. Sc) spirals are pseudo-bulges, the difference
of stellar population between early and late type spirals will be even larger
because the pseudo-bulges are bluer than classical bulges. This will increase
the predicted $\Delta M$ between the early and late type spirals and decrease
the allowed differences between their disks again.

\subsection{dust extinction}\label{dust}

In our modeling,  we have not considered the  effect of internal dust
extinction, which might be different for different type spirals.

Fortunately, the internal dust extinction has at least been partly corrected in
most of the observational TFR studies, e.g.
\citet{Giovanelli97,Masters06,Masters08}. In these studies, they all made the
corrections of the internal dust extinction using the parametrization of
\citet{Giovanelli94},

\begin{equation}\label{dM}
A =\gamma\,{\rm{Log}}(a/b)\,
\end{equation}
where $\gamma$ is a waveband dependent coefficient and $a/b$ is the observed
axis ratio indicating the inclination angle of the spiral disk. This
parametrization of the dust extinction has accounted  the extinction from the
geometry effect, i.e. it has corrected all the spiral disk to face-on viewing
and assumed zero dust extinction for face-on case($A\sim 0$ for $a/b\sim 1$).

However, the internal dust extinction of face-on spirals may not be negligible
\citep{Shao07,Driver07}. The inner parts of the spiral disks are very likely to
be optical thick even when viewed in face-on\citep{Giovanelli94}. The
dependence of this face-on extinction on the morphological type is complicate.
On  one hand, the later type spiral disks are gas richer and so that may
contain more dust \citep{Stevens05}. On the other hand, the earlier type
spirals have larger bulge component and this component suffers more from the
dust extinction because the dust layer is peaked in the central region
\citep{Tuffs04,Driver07}. Counteracting these two effects, the global dust
extinction may only be weakly dependent on the morphological types. A very
recent study of \citet{Munoz09} based on the multi-wavelength data shows that
there is a weak global trend of the internal dust extinction, Sb$>$Sa$>$Sc.

If the face-on dust extinction of early type spirals were more significant
considering their more concentrated light, the smaller $\Delta M$ predicted by
the `composition' model in $B$ band could  be explained since the extinction in
$B$ band is  5 times more larger than $K$ band given the typical extinction
curve of normal spiral galaxies\citep{Shao07}. Moreover, if we took the higher
dust extinction for the early type spirals into our model, the model predicted
$\Delta M$ would be  larger, thus the stellar populations of different type
disks even could not be very  different as that suggested by K94.

\section{summary}
In this paper, we have studied the morphological dependent TFRs by modeling the
dynamics and stellar population of spiral galaxies in combination. We model the
dynamics of  spiral galaxies by following the disk formation model of MMW and
extending it to include a bulge component. We parameterize the SFHs of the
bulge and disk separately and derive their mass-to-light ratios  with
population synthesis code of \cite{BC03}. Our model reproduces the observed $I$
band TFR of the Sc spirals very well without any free parameters. Our model
shows that the morphological dependence of the TFRs is mainly contributed by
the stellar population through bulge disk composition effect, although  the
effect from the dynamics is not negligible.

The effects of the dynamics and stellar population on the morphological
dependence of the TFR are different.  The  shift of the $\Vmax$ (or
luminosity), caused by dynamics, shows a characteristic that is independent of
the wavebands. On the other hand, the shift of the zero-point of TFR is a
function of the wavebands, when the morphological dependence is caused by the
stellar population. Therefore, the morphological dependence of the TFR in
different bands is  an effective way to constrain the dynamics and stellar
populations of spiral galaxies. Our results show that the observed
morphological dependent TFRs are consistent with a scenario that the stellar
population of different type spirals are only different in the bulge disk
composition on average.

\section*{Acknowledgments}
We thank the anonymous referee for posing questions which significantly
clarified the analysis in this paper. This project is supported by
NSFC10803016, NSFC10833005,NKBRSF2007CB815402, Shanghai Rising-Star
Program(08QA14077) and Shanghai Municipal Science and Technology Commission No.
04dz\_05905.

\newpage
\begin{longtable}{lcccccccccc}

\caption{The observed morphological dependent TFRs in literatures. All the TFRs
are re-shaped in the form of Equ. (1). $\Delta M$ is the shift of the zero-point $\beta$ relative to the Sc spirals in the
same study.}
\endhead

 \hline
   Reference & Waveband  & type  & sub-type   &   $a$  &   $b$ & $\Delta M$  \\
 \hline
   \cite{Sandage00}, Table 13 &   $B$       &  Sa   & Sa/Sab     & $-6.97$ & $-20.46\pm0.12$ &  $0.68\pm.13$\\
                              &             &  Sc   & Sbc/Sc/Scd & $-6.97$ & $-21.14\pm0.06$ &  0 \\
\hline
     \cite{Russell04} Equ. 2 and 3 &  $B$    &  Sb   & Sab/Sb/Sbc/Sc II-IV & $-5.24\pm0.10$ & $-20.42\pm0.16$ & $0.51\pm.18$ \\
	                          &          &  Sc   & Sbc/Sc I-II & $-4.91\pm0.20$ & $-20.93\pm0.09$ & 0 \\
\hline
\cite{Masters06}, Table 4    & $I$      &  Sa   & S0/Sa/Sab     & $-5.52\pm0.40$ & $-21.34\pm0.07$ & $0.32\pm0.07$ \\
                             &          &  Sb   & Sb            & $-7.07\pm0.17$ & $-21.45\pm0.03$ & $0.21\pm0.08$  \\
                             &          &  Sc   & Sbc/Sc/Scd    & $-7.87\pm0.15$ & $-21.66\pm0.02$ &  0 \\
\hline
  \cite{Masters08}, Table 2 & $J$ & Sa & S0/Sa/Sab & $-6.09\pm0.30$ & $-21.84\pm0.06$ & $0.12\pm0.06$\\
                            &     & Sb & Sb/Sbc    & $-7.80\pm0.19$ & $-21.81\pm0.03$ & $0.15\pm0.04$\\
                            &     & Sc & Sc/Scd    & $-9.23\pm0.20$ & $-21.96\pm0.02$ &     0        \\
                            & $H$ & Sa & S0/Sa/Sab & $-6.08\pm0.27$ & $-22.62\pm0.05$ & $0.16\pm0.05$\\
                            &     & Sb & Sb/Sbc    & $-7.80\pm0.19$ & $-22.64\pm0.03$ & $0.14\pm0.04$\\
                            &     & Sc & Sc/Scd    & $-9.17\pm0.20$ & $-23.78\pm0.02$ &         0 \\
                            & $K$ & Sa & S0/Sa/Sab & $-6.95\pm0.24$ & $-22.85\pm0.03$ & $0.21\pm0.04$\\
                            &     & Sb & Sb/Sbc    & $-8.64\pm0.17$ & $-22.92\pm0.02$ & $0.14\pm0.04$\\
                            &     & Sc & Sc/Scd    & $-10.09\pm0.18$ & $-23.06\pm0.02$ & 0          \\
\hline
 \cite{Russell08}            & $K_s$  &Sb & Sa/Sab/Sb & $-$ & $-$                      & $0.19\pm0.10$ \\
\hline

\end{longtable}

\end{document}